\documentclass{ws-mpla}
\usepackage[super,compress]{cite}
\usepackage{graphicx}
\usepackage{hyperref}
\hypersetup{colorlinks,urlcolor=blue,citecolor=black,linkcolor=black,filecolor=black}
\usepackage{color}
\usepackage{orcidlink}

\begin{document}
\markboth{F. Ahmed}{five-dimensional vacuum-defect wormhole 
}

\catchline{}{}{}{}{}

\title{\bf A five-dimensional vacuum-defect wormhole space-time
}

\author{Faizuddin Ahmed\orcidlink{0000-0003-2196-9622}
}
\address{Department of Physics, University of Science \& Technology Meghalaya, Ri-Bhoi, Meghalaya, 793101, India\\
E-mail: {\tt faizuddinahmed15@gmail.com}
}

\maketitle

\begin{history}
\received{Day Month Year}
\revised{Day Month Year}
\end{history}

\begin{abstract}
In this study, we present a novel extension to the Klinkhamer vacuum-defect model by incorporating a fifth spatial dimension. This modification results in the formulation of a comprehensive five-dimensional wormhole space-time. Crucially, this extension preserves its classification as a vacuum solution to the field equations, thereby automatically satisfying all energy conditions. We demonstrate that this five-dimensional vacuum-defect space-time can be expressed in a pure canonical form, and thus, locally reduces to five-dimensional Minkowski space. Finally, we explore the geodesic equations in the vicinity of this five-dimensional vacuum-defect wormhole, offering insights into its intriguing properties.


\end{abstract}

\section{Introduction}

Einstein's four-dimensional theory of general relativity was extended to five dimensions by Kaluza and Klein \cite{ss6,ss7} in an effort to unify gravitation and electromagnetism. This innovative concept proposed that electromagnetism could be incorporated by introducing an additional, compactified dimension within the framework of four-dimensional space-time. In this model, the formerly spatial fifth dimension became an integral part of the unified theory.

The five-dimensional model employs a metric tensor denoted as $g_{AB}$, with a corresponding line element given by $dS^2 = g_{AB}\,dx^{A}\,dx^{B}$, where the indices $A, B$ run through values $0, 123, 4$ to represent time, space, and the extra dimension. This framework encompasses a four-dimensional manifold characterized by a line element $ds^2 = g_{\mu\nu}\,dx^{\mu}\, dx^{\nu}$, where indices $\mu, \nu$ take on values $0, 123$ to denote the four-dimensional space-time. Within this context, the null-geodesics in the five-dimensional space-time, given by $dS^2 = 0$, describe the motion of massive objects in the four-dimensional space-time, where $ds^2 < 0$ \cite{ss1,ss2,ss3}. This connection between null-geodesics and massive object motion necessitates the introduction of a relation such as fifth coordinate $x^4 =\chi=\chi(s)$, involving the proper time in the four-dimensional space-time, to account for the embedding of the four-dimensional manifold into the five-dimensional case. When the four-dimensional component $g_{\mu\nu}$ within the five-dimensional space-time is explicitly independent of the fifth dimension, specifically $g_{\mu\nu}=g_{\mu\nu}(x^{\alpha})$ and $\partial g_{\mu\nu}/\partial \chi = 0$, the fifth coordinate assumes the role of particle mass, and this formulation adheres to the weak equivalence principle \cite{ss3,ss4}. This also implies that the fifth force is absent, leaving the accelerations exactly as they are in four-dimensional theory. Null geodesics in five-dimensional manifolds has been studied in \cite{ss1}.  

The vacuum Einstein's field equations for space-time plus extra dimension are given in terms of the Ricci tensor by $R_{AB}=0$ (where $A,B=0, 123, 4$). This is the simplest type of higher-dimensional field equations in general relativity. This condition is relevant to the space-time-matter (STM), but can be applied to some other scenarios. There are a number of known solutions that embed four-dimensional manifolds with or without cosmological constant and/or spherically symmetric character \cite{ss8}, whereby the extended version is known to agree with observation \cite{ss8,ss9,ss10}. A few examples are static wormhole solutions within the framework of five-dimensions Kaluza-Klein gravity in the presence of a massless ghost four-dimensional scalar field \cite{ss11}, five-dimensional Einstein equations with four-dimensional de Sitter-like expansion \cite{ss12}, five-dimensional G\"{o}del-type space-time \cite{ss13,ss15}, FRW cosmological model \cite{ss14}, cosmological implications of nonseparable five-dimensional vacuum field equations \cite{ss16}, spherically symmetric dyonic wormhole-like space-time in five-dimensional Kaluza-Klein theory \cite{ss17}, axisymmetric regular multiwormhole \cite{ss20} and stationary solutions \cite{ss18,ss19} in five-dimensional general relativity theory, wormholes in $(4+1)$ gravity \cite{ss21}, a regular vacuum solution in 5D gravity \cite{ss22}, classical and quantized aspects of dynamics in five-dimensional relativity \cite{ss3}, and five-dimensional black holes \cite{ss24}.

In searching for a new five-dimensional solution to vacuum field equations, we must keep in mind a known vacuum solution in four-dimensional theory. Our investigation is based on a four-dimensional vacuum-defect wormhole metric presented in \cite{FRK2} given by the following line-element (see also related work \cite{FRK})
\begin{equation}
    ds^2|^{4D}_{\mbox{vacuum-defect}}=-dt^2+\Big(1+\frac{b^2}{\xi^2}\Big)^{-1}\,d\xi^2+(\xi^2+b^2)\,(d\theta^2+\sin^2 \theta\,d\phi^2),
    \label{K1}
\end{equation}
where $-\infty < (t, \xi) < +\infty$ and other coordinates are in the usual ranges. This four-dimensional defect wormhole metric is not only Ricci flat, $R_{\mu\nu}=0$ but also the Riemann flat, $R_{\mu\nu\alpha\beta}=0$. The $4D$ defect wormhole has investigated in modified gravity, such as Schwarzschild-Klinkhamer wormhole with global monopole charge \cite{FA}, topologically charged wormhole supporting the energy condition \cite{FA2}, and topologically charged rotating wormhole \cite{FA3}. It is worth noting that in the context of vacuum-defect wormhole, Feng \cite{CQG} and Visser {\it et al.} \cite{Uni} independently put forwarded their comments and discussion. 

Our primary objective in this study is to construct a novel five-dimensional vacuum-defect wormhole model, which serves as a natural extension or generalization of the existing vacuum-defect wormhole (\ref{K1}). Particular interest given around a solution which are not only the Ricci flat, $R_{AB}=0$ but also Riemann flat, that is, the curvature tensor is $R_{ABCD}=0$ in five-dimensions. Notably, we focus on maintaining the vacuum nature of the five-dimensional solution by excluding any incorporation of a mass parameter, $M$, in contrast to the approach taken in the work \cite{FRK3}, where a higher-dimensional extension of the vacuum-defect wormhole was achieved by introducing a mass parameter $M$ linked to the standard $4D$ Schwarzschild metric.

In this research endeavor, we meticulously solve the Einstein's field equations to derive a five-dimensional vacuum-defect wormhole metric. Importantly, this five-dimensional wormhole metric seamlessly extends from the four-dimensional vacuum-defect wormhole metric (\ref{K1}), preserving its essential attributes while encompassing an additional spatial dimension, as fifth dimensions. Our subsequent analysis focuses on comprehensively examining the geodesic motion of test particles surrounding this novel five-dimensional vacuum wormhole.

{\it Conventions.} Throughout the paper, the $5D$ metric signature is taken to be $(+, -, -, -, -)$, while the choice of $4D$ metric signature is $(+,-,-,-)$. The space-time coordinates are labeled as $x^0=t$, $x^i=(\xi, \theta, \phi)$. The extra coordinate is $x^4=\chi$. The range of Latin letters are as follows: $A,B,...=0-4$ and the Greek indices $\mu,\nu=0-3$. Finally, we use geometric units where $c=1=8\,\pi\,G=\hbar$.

\section{Analysis of a five-dimensional vacuum-defect wormhole}

As stated above, we aim to extend the four-dimensional Klinkhamer-vacuum defect wormhole (\ref{K1}) into the higher-dimensional theory, especially in the five-dimensional manifold. Therefore, the line-element describing this five-dimensional wormhole metric in the chart $(t, \xi, \theta, \phi, \chi)$ is given by
\begin{eqnarray}
    ds^2|^{5D}=ds^2|^{4D}_{\mbox{vacuum-defect}}+F(\xi,\theta)\,d\chi^2,
    \label{1}
\end{eqnarray}
where the four-dimensional part $ds^2|^{4D}_{\mbox{vacuum-defect}}$ is given by (\ref{K1}), $\chi$ is the fifth coordinate of extra dimensions which is a spatial with range $\chi \in (0, \infty)$, and the function $F(\xi, \theta)=D(\xi)\,H(\theta)$ with $D(\xi)$ and $H(\theta)$ are unknown functions that are to be determined.

By choosing this particular form of the $5D$ space-time, we aim to construct the $5D$ vacuum field equations given by
\begin{equation}
    G_{AB}=0\quad\quad \mbox{or}\quad\quad R_{AB}=0.\label{ricci}
\end{equation}
We will now show that given a metric of the form (\ref{1}) and the field equations (\ref{ricci}), then the function $F(\xi,\theta)$
is determined in terms of $\xi$ and $\theta$.

For the five-dimensional metric (\ref{1}), the Einstein tensor $G_{AA} \neq 0$ along with a non-zero non-diagonal term given by 
\begin{eqnarray}
     G_{\theta\xi}=G_{\xi\theta}=\frac{H'}{4\,H}\,\Big(\frac{2\,\xi}{\xi^2+b^2}-\frac{D'}{D}\Big),
    \label{3}
\end{eqnarray}
where $H'$ and $D'$ denotes derivative w. r. t. $\theta$ and $\xi$, respectively. 

Solving for vacuum field equations and considering first $G_{\theta\xi}=0$, from equation (\ref{3}), we obtain
\begin{equation}
    D(\xi)=c_1\,(\xi^2+b^2),
    \label{4}
\end{equation}
where $c_1>0$ is an arbitrary constant choosing unity ($c_1=1$) here for simplicity.

With this function $D(\xi)$ given in Eq. (\ref{4}), the five-dimensional line-element (\ref{1}) now can be written as
\begin{eqnarray}
    ds^2=-dt^2+\Big(1+\frac{b^2}{\xi^2}\Big)^{-1}\,d\xi^2+(\xi^2+b^2)\,\Big(d\theta^2+\sin^2 \theta\,d\phi^2+H(\theta)\,d\chi^2\Big),
    \label{5}
\end{eqnarray}
Still space-time (\ref{5}) is a non-vacuum solution of the field equations, $G_{AB} \neq 0$. The non-zero components of the Einstein tensor $G_{AB}$ for this metric (\ref{5}) are given by 
\begin{eqnarray}
    &&G_{tt}=R/2,\nonumber\\
    &&G_{\xi\xi}=-\frac{\xi^2\,R}{2},\nonumber\\
    &&G_{\theta\theta}=1+\frac{H'}{2\,H}\,\cot \theta,\nonumber\\
    &&G_{\phi\phi}=\frac{\sin^2 \theta}{4\,H}\,\Big(-H'^2+4\,H^2+2\,H\,H''\Big).
    \label{6}
\end{eqnarray}
where $R$ is the Ricci scalar given by
\begin{eqnarray}
   R=\frac{-8\,H^2+H'^2-2\,H\,(H'\,\cot \theta+H'')}{2\,(\xi^2+b^2)\,H^2}.
   \label{7}
\end{eqnarray}

For vacuum field equations, the Einstein tensor, $G_{AB}=0$, and the Ricci scalar. $R=0$. From equations (\ref{6}) and (\ref{7}), we obtain the following set of equation given by
\begin{eqnarray}
    &&1+\frac{H'}{2\,H}\,\cot \theta=0,\nonumber\\
    &&H'^2=4\,H^2+2\,H\,H'',\nonumber\\
    &&H'^2=8\,H^2+2\,H\,(H'\,\cot \theta+H'').
    \label{8}
\end{eqnarray}
Simplification of the above set of equation result the function $H(\theta)$ to be the following form
\begin{equation}
    H(\theta)=c_2\,\cos^2 \theta,
    \label{8a}
\end{equation}
where $c_2>0$ is another constant choosing unity here for simplicity. 

Therefore, the final form of the five-dimensional vacuum wormhole space-time (\ref{5}) using the function (\ref{8a}) is of the following form
\begin{eqnarray}
    ds^2|^{5D}&=&-dt^2+\Big(1+\frac{b^2}{\xi^2}\Big)^{-1}d\xi^2+(\xi^2+b^2)\,\big(d\theta^2+\sin^2 \theta\,d\phi^2\big)\nonumber\\
    &+&(\xi^2+b^2)\,\cos^2 \theta\,d\chi^2.
    \label{9}
\end{eqnarray}

We see that the $4D$ part of the $5D$ metric tensor $g_{AB}$ doesn't depend on the extra coordinate $\chi$, and hence, $\partial g_{\mu\nu}/\partial \chi=0$, where $g_{\mu\nu}$ is the metric tensor of the $4D$ manifold given by (\ref{K1}). Thus, the metric (\ref{9}) is in fact a pure-canonical form in $5D$ and is called canonical metric or the warp metric denoted by $C_5$ \cite{qq7,qq8,qq1}. A few other known vacuum solutions of the $5D$ field equations which are canonical type in form are reported in \cite{qq1,qq2,qq3,qq4,qq5,qq6}. 

One can calculate the Kretschmann scalar curvature for this five-dimensional vacuum-defect wormhole  space-time (\ref{9}) and it will be zero, $\mathcal{K}=R^{ABCD}\,R_{ABCD}=0$. In addition, this metric (\ref{9}) not only Ricci-flat, $R_{AB}=0$, but also the Riemann-flat since the Riemann-Christoffel tensor is $R_{ABCD}=0$ (vanishing of the Riemann-Christoffel tensor means that we are considering the analogue of the Minkowski metric in 5D) \cite{ss10,qq5,qq6,qq9,qq10,qq11}. Hence, there is a set of coordinates transformations for which the metric (\ref{9}) becomes locally five-dimensional Minkowski space ( denoted as $M_5$) given by
\begin{equation}
    ds^2|^{5D}_{\mbox{Minkowski}}=-dT^2+dX^2+dY^2+dZ^2+dW^2.
    \label{9a}
\end{equation}
The precise transformations between $(T, X, Y, Z, W)$ and $(t, \xi, \theta, \phi,\chi)$ are given by $T \to t$ and
\begin{eqnarray}
    &&X=\sqrt{\xi^2+b^2}\,\sin \theta\,\cos \phi,\quad Y=\sqrt{\xi^2+b^2}\,\sin \theta\,\sin \phi,\nonumber\\
    &&Z=\sqrt{\xi^2+b^2}\,\cos \theta\,\cos \chi,\quad W=\sqrt{\xi^2+b^2}\,\cos \theta\,\sin \chi.\label{9b}
\end{eqnarray}

Moreover, in four-dimensional cases such as for the metric (\ref{K1}), we have the transformation $T \to t$ and
\begin{equation}
    X=\sqrt{\xi^2+b^2}\,\sin \theta\,\cos \phi,\quad Y=\sqrt{\xi^2+b^2}\,\sin \theta\,\sin \phi,\quad Z=\sqrt{\xi^2+b^2}\,\cos \theta
    \label{9c}
\end{equation}
that gives us the $4D$ Minkowski space $M_4$ as follows: 
\begin{equation}
    ds^2|^{4D}_{\mbox{Minkowski}}=-dT^2+dX^2+dY^2+dZ^2.
    \label{Min}
\end{equation}

We see that this five-dimensional wormhole space-times (\ref{9}) admits 15 killing vector fields. The norm of the killing vector field $X_{\chi}=\partial_{\chi}$ given by
\begin{equation}
    X^2_{\chi}=|\partial_{\chi} \bullet \partial_{\chi}|=g_{\chi\chi}=(\xi^2+b^2)\,\cos^2 \theta\,\, \geq\,\, 0
    \label{killing}
\end{equation}
is spacelike (or null) and finite at $\xi=0$ as well as on the axis of rotation $\theta=0, \pi$. Furthermore, we also see that for $\chi=const$ hypersurface, the wormhole space-time (\ref{9}) reduces to the vacuum-defect defect metric (\ref{K1}). That means the four-dimensional Klinkhamer vacuum-defect wormhole is embedded in a five-dimensional space $M_5$ with the metric given by (\ref{9}).

Now, we study the geodesics motions of test particles around this new five-dimensional vacuum wormhole space-time (\ref{9}). The geodesic equations are given by
\begin{eqnarray}
    &&\ddot{t}=0,\\
    \label{11}
    &&\ddot{\xi}=\frac{b^2}{\xi\,(b^2+\xi^2)}\,\dot{\xi}^2+\frac{\xi^2+b^2}{\xi}\,\dot{\theta}^2+\frac{(\xi^2+b^2)\,\sin^2 \theta}{\xi}\,\dot{\phi}^2+\frac{(\xi^2+b^2)\,\cos^2 \theta}{\xi}\,\dot{\chi}^2,\\
    \label{12}
    &&\ddot{\theta}=\frac{2\,\xi}{\xi^2+b^2}\,\dot{\xi}\,\dot{\theta}+\frac{1}{2}\,\sin 2\theta\,(\dot{\phi}^2-\dot{\chi}^2),\\
    \label{13}
    &&\ddot{\phi}=-\frac{2\,\xi}{\xi^2+b^2}\,\dot{\xi}\,\dot{\phi}-2\,\cot \theta\,\dot{\theta}\,\dot{\phi},\\
    \label{14}
    &&\ddot{\chi}=2\,\Big(-\frac{\xi}{\xi^2+b^2}\,\dot{\xi}+\tan \theta\,\dot{\theta}\Big)\,\dot{\chi}.
    \label{15}
\end{eqnarray}
where dot represents derivative w. r. t. affine parameter $\tau$.

The first derivative of last two equations results
\begin{eqnarray}
    &&\dot{\phi} (\tau)=\frac{c_3}{(\xi^2+b^2)\,\sin^2 \theta},\nonumber\\
    &&\dot{\chi} (\tau)=\frac{c_4}{(\xi^2+b^2)\,\cos^2 \theta},
    \label{16}
\end{eqnarray}
where $c_3, c_4$ are arbitrary positive constants. Substituting Eq. (\ref{16}) into the equations (\ref{11})-(\ref{12}), one will get the second-order equations of $\xi, \theta$ whose solutions are little bit complicated. 

On the other hand, in the particular case, where $\theta=const \neq \pi/2$, the geodesics equations from equations (\ref{11})-(\ref{15}) reduces to as follows:
\begin{eqnarray}
    &&\dot{t} (\tau)=const,\nonumber\\
    &&\dot{\phi}(\tau)=\frac{c_0}{\xi^2+b^2}=\dot{\chi}(\tau),\nonumber\\ 
    &&\ddot{\xi}=\frac{b^2\,\dot{\xi}^2+c^2}{\xi\,(\xi^2+b^2)}.
    \label{17}
\end{eqnarray}
where $c_0$ is an arbitrary constant. Thus, we see that the geodesics path for the fifth coordinate $\chi$ behaves the same character as the azimuthal coordinate $\phi$ in the plane defined by $\theta=const$.

The null-path in five-dimensions with $ds^2|^{5D}=0$ in the metric (\ref{9}) corresponds to the time-like path in four-dimensions with $ds^2|^{4D}=-(\xi^2+b^2)\,\cos^2\theta\,d\chi^2 <0$ of a massive object \cite{ss1,ss2,ss3}. This means that a massive object or particle in the four-dimensional metric behaves as photon-like in the five-dimensional metric. In other words, all massive objects in the four-dimensional metric travel along null paths in the five-dimensional metric. 

\section{Conclusions}   

In this work, we presented a five-dimensional vacuum-defect wormhole metric serves as an extension of the established four-dimensional Klinkhamer-vacuum defect model. It's worth highlighting that upon scrutinizing the geometric characteristics of both vacuum-defect wormhole models, as denoted by equations (\ref{K1}) and (\ref{9}), a striking similarity becomes evident, particularly when considering in the equatorial plane defined by $\theta=\pi/2$. Furthermore, an intriguing observation arises when investigating our wormhole model on surfaces where $\chi$ remains constant. One can see that the well-established Klinkhamer-vacuum defect metric is embedded in the five-dimensional wormhole metric. A major result is that null geodesics in $5D$ can correspond to non-null geodesics in $4D$. That is, massless particles in $4D$ can correspond to massive particles in $5D$. In the future studies, we will focus on constructing a $5D$ vacuum-defect wormhole with a cosmological constant and analyze the result.

\section*{Acknowledgement}

We sincerely acknowledges the anonymous referees for their valuable comments and helpful suggestions. F.A. acknowledges the Inter University Centre for Astronomy and Astrophysics (IUCAA), Pune, India for granted visiting associateship.

\section*{Data Availability Statement}

No new data were generated or analyzed in this paper.

\section*{Conflict of Interest}

There is no such conflicts of interests.

\end{document}